\documentclass[prb,twocolumn,aps,superscriptaddress]{revtex4-2}
\usepackage{graphicx}
\usepackage{dcolumn}
\usepackage{bm}
\usepackage{amssymb}
\usepackage{amsmath}
\usepackage{physics}
\usepackage[utf8]{inputenc}

\usepackage{hyperref}
\usepackage{graphicx}
\usepackage{xcolor}
\usepackage{tabularx}
\usepackage{float}
\floatstyle{plaintop}
\restylefloat{table}

\usepackage{soul}
\hypersetup{colorlinks = true, linkcolor=blue, citecolor=blue, urlcolor=}
\usepackage[T1]{fontenc}

\begin{document}


\title{Steady-state dynamics and nonlocal correlations in thermoelectric Cooper pair splitters}
\author{Arnav Arora}
\email{denotes equal contribution}
 \affiliation{Department of Physics, Indian Institute of Technology Roorkee, India
}

\author{Siddhant Midha}
\email{denotes equal contribution}
 \affiliation{Department of Electrical Engineering, Indian Institute of Technology Bombay, Powai, Mumbai--400076, India
}
\author{Alexander Zyuzin}
 \affiliation{Low Temperature Laboratory, Department of Applied Physics, Aalto University, PO Box 15100, FI-00076, Espoo, Finland
}
 \affiliation{QTF Centre of Excellence, Department of Applied Physics, Aalto University, PO Box 15100, FI-00076, Espoo, Finland
}
\author{Pertti Hakonen}
 \affiliation{Low Temperature Laboratory, Department of Applied Physics, Aalto University, PO Box 15100, FI-00076, Espoo, Finland
}
 \affiliation{QTF Centre of Excellence, Department of Applied Physics, Aalto University, PO Box 15100, FI-00076, Espoo, Finland
}
\author{Bhaskaran Muralidharan}
\email{bm@ee.iitb.ac.in}
 \affiliation{Department of Electrical Engineering, Indian Institute of Technology Bombay, Powai, Mumbai--400076, India
}
\affiliation{Center of Excellence in Quantum Information, Computing Science and Technology, Indian Institute of Technology Bombay, Powai, Mumbai--400076, India}

\date{\today}
\begin{abstract} Recent experiments on Cooper pair splitters using superconductor-quantum dot hybrids have embarked on creating entanglement in the solid-state, by engineering the sub-gap processes in the superconducting region. Using the thermoelectric Cooper pair splitter setup [Nat. Comm., 12, 21, (2021)] as a prototype, we present a comprehensive analysis of the fundamental components of the observed transport signal, aiming to critically clarify the operating regimes and confirm the nonlocal and nonclassical nature of correlations arising from crossed Andreev processes. By making a nexus with quantum discord, we identify operating points of nonlocal quantum correlations in the CPS device---information that cannot be extracted from the transport signal alone. A notable consequence of our analysis is the finding that contact-induced level broadening of the quantum dot's discrete energy spectrum, along with its hybridization with the superconducting segment, can lead to shifted resonances in the crossed Andreev process as well as a parity reversal in the thermoelectric current. Our work thereby provides detailed insights into the gate voltage control of the quantum correlations in superconducting-hybrid Cooper pair splitters, revealing new avenues for harnessing quantum correlations in solid-state systems.
\end{abstract}

\maketitle

\section{Introduction}
Leveraging the coupling between the superconducting ground state and spatially separated quantum dots, Cooper pair splitters (CPS) have emerged as viable candidates for generating entanglement in solid-state systems \cite{electronicentanglement1,electronicentanglement2}. These devices employ the phenomenon of crossed Andreev reflection (CAR) \cite{CAR}, a subgap transport process that occurs between two distinct superconductor-normal (SN) interfaces, to create nonlocal spin-entangled electrons. Recent experiments have reported progress in realizing CPS in numerous systems, including carbon nanotubes \cite{cnt-exp-1, cnt-exp-2}, SN heterostructures \cite{metallic-island-exp-1, metallic-island-exp-2}, semiconducting nanowires \cite{nw-exp-1, nw-exp-2, nw-exp-3, nw-exp-4, nw-exp-5}, semiconducting \cite{semi-qd-exp-1, semi-qd-exp-2, semi-qd-exp-3} and graphene quantum dots \cite{graphene-exp-1, graphene-exp-3-sqds, graphene-exp-4}, to name a few. State-of-the-art techniques such as charge measurement \cite{metallic-island-exp-1, metallic-island-exp-2}, microwave readout \cite{semi-qd-exp-3} and thermoelectric measurements \cite{Splitter_Hakonen,Hakonen_PRB,Cao_APL,Hakonen_PRL,Sanchez_Cooling} are employed to probe nonlocal currents arising in CPS devices \cite{cps_expprobing1,cps_expprobing2,cps_expprobing3,cps_expprobing4,cps_expprobing5,cps_expprobing6,cpswitnessing}. \\
\begin{figure*}[t]
    \centering
     \includegraphics[height=8cm, width = \textwidth]{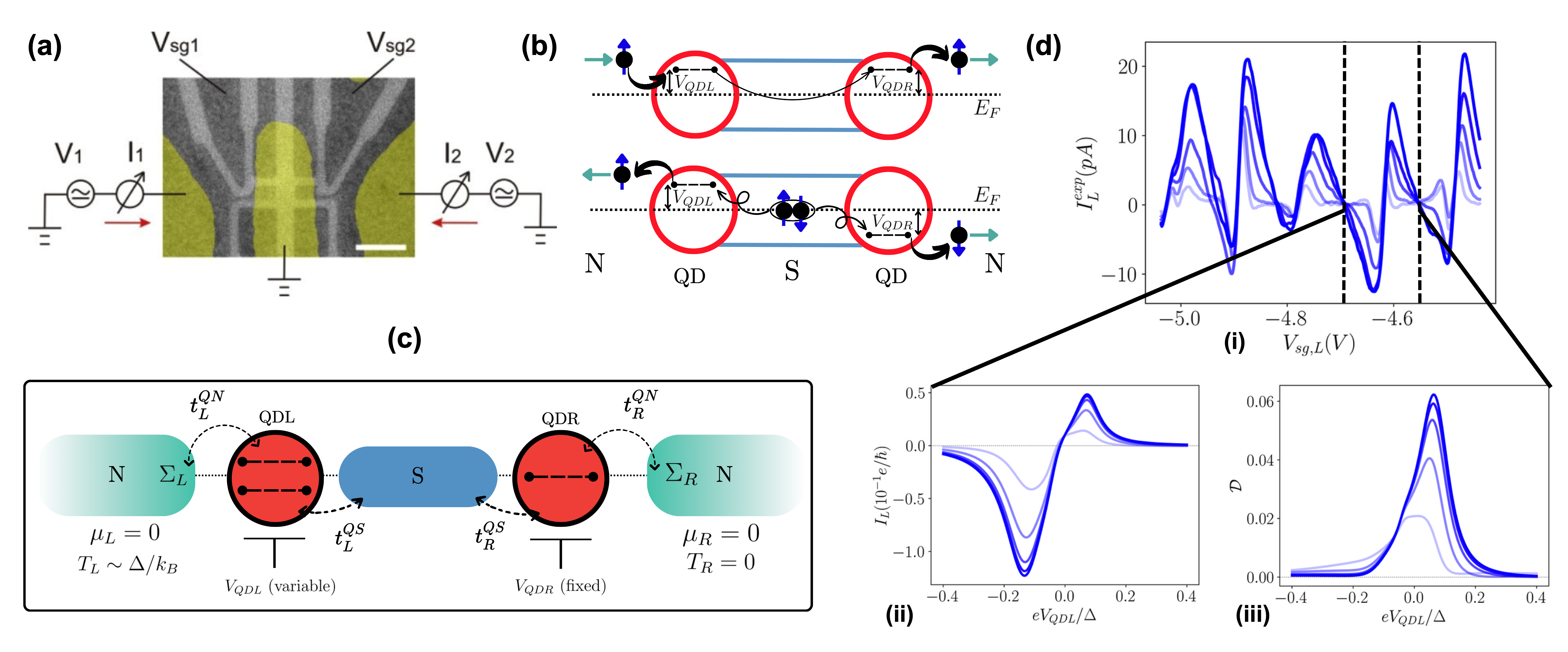}
    \caption{{\bf{Thermoelectric Cooper Pair Splitter}} (a) False colour image of the QD-CPS hybrid device shown to exhibit nonlocal currents under a thermoelectric bias \cite{Splitter_Hakonen}, with the heater situated on the left side (not shown here). $V_1$, $V_2$, $I_1$, $I_2$ represent the left and right contact voltmeter and ammeters respectively. $V_{sg1}$ and $V_{sg2}$ are the left and right QD gate-voltages. (b) A schematic of the ECT (top) and CAR (bottom) mechanisms in the case of un-hybridized QDs, with ECT dominating at the $(E,E)$ resonance and CAR dominating at the $(E,-E)$ resonance. $E_F$ denotes the Fermi level of the system. (c) A detailed schematic of the NEGF-based theoretical model used in this paper consisting of a QD-SC-QD Hamiltonian perturbed by two normal contacts at zero voltage bias potential $\mu_L = \mu_R = 0$, with QD-N coupling denoted $t^{QN}_{L/R}$ and QD-S coupling denoted $t^{QS}_{L/R}$. The left contact is placed under a thermal bias $\delta T := (T_L - T_R)$ of order $k_B\delta T\sim\Delta$. The left-dot voltage is varied for analysis. (d) Analyzing currents through the device as a function of thermal bias. Panel (i) shows the experimental local TE current ($I^{exp}_L$ in pico-Amperes) sweep through the QDL as a function of the left gate bias ($V_{sg1}$ corresponding to (c) in Volts) revealing a series of bi-lobe (sawtooth like) current structures for heater voltages $V_h \in \{5, 9, 19, 25, 29\}$mV whereby the temperature depends on the heater voltage as $T \propto V^{0.7}$(see\cite{Splitter_Hakonen}). Panel (ii)
    shows a zoomed-in view of a single current ($I_L$ in units of $10^{-1}e/\hbar$) bi-lobe structure from our theoretical calculations for $k_B\delta T \in \{0.04,0.07,0.1,0.2,0.3\}$. Panel (iii) shows the quantum discord $\mathcal{D}$ between the two QDs computed as a function of $V_{QDL}$. The color code represents darker colors for increasing temperatures in (i-iii).}
    \label{fig:fig1}
\end{figure*}
\indent Coupling a superconducting (SC) region with quantum dots (QDs) at its ends permits the spectral probing of the subgap processes, in which the well-spaced and unbroadened QD levels can act as probes facilitated by electrostatic gating. In addition to the CAR processes, such devices also exhibit competing elastic co-tunneling (ECT) processes between the two QDs, as well as local Andreev reflections (LAR) at each QD~\cite{CPSold}. In particular, a thermoelectric setup eliminates local Andreev processes \cite{thermo-theo-3} which are generally orders of magnitude larger than nonlocal processes that contribute to the CPS signal. Apart from the CPS configuration, this setup provides a framework for the subgap engineering of CAR and ECT processes in connection with the minimal Kitaev chains for the realization of poor man's Majoranas \cite{Aguado_2024,PRXQuantum.5.010323,PMM_Klino}. In this paper, we theoretically advance the interpretation of the thermoelectric CPS experiment \cite{Splitter_Hakonen}, not only by giving a detailed explanation of the observed currents, but also by providing new means to establish the nonlocality of the obtained CPS transport signal. \\
\indent Using the experimental setup  \cite{Splitter_Hakonen} depicted in Fig. \ref{fig:fig1}(a) as our prototype, we employ the Keldysh non-equilibrium Green's function (NEGF) framework to unravel new insights into the correlations and transport signals generated in the CPS device. Focusing on the thermoelectric (TE) signal noted in the experiment, we investigate in detail the presence of nonlocal quantum correlations arising at the Cooper pair splitting resonance. Setting up the NEGF approach for the CPS with minimal assumptions, we resolve the currents into the ECT and CAR components, provide vital insights into the spectral structure of the currents and the QD-SC density of states (DOS). We also compute the quantum discord \cite{fermionicdiscord,quantumdiscord} (see Sect. \ref{METH}) between the spatially separated QDs to thereby enable us to pin-point the occurrence of nonlocal correlations that are only CAR induced.  \\  
\indent Apart from the NEGF technique \cite{CPS_Keldysh1,CPS_Keldysh2,CPS_Keldysh3,CPS_Keldysh4}, prior theoretical approaches to analyze CPS have been based on semi-classical rate equations \cite{Palyi-semi-classical,buttiker-semi-classical}, quantum master equations \cite{cps_flindt_adiabatic,cps_flindt_dynamics,cps_thermotheory,martin_master_eq,cps_flindt_entwitness,cps_flindt_countingstats}, or using a transmission formalism across individual segments \cite{cps_thermotheory,Splitter_Hakonen} of the device. The master equation approaches are, no doubt, pertinent since the QDs are typically in the Coulomb blockade (CB) regime, such that the charging effects are accounted for within the Fock-space \cite{Das_2022} of the sub-system. However, in such approaches, the processes of CAR and ECT are accounted for by describing the SC region via effective coupling parameters. 
Furthermore, in the present experiment \cite{Splitter_Hakonen}, we will demonstrate that understanding the discreteness of the energy levels, the hybridization effects between the quantum dots, superconducting regions, and contacts, as well as the resulting energy level broadening, is essential to fully explain the observed transport signatures.  \\
\indent Our approach enables a comprehensive insight into the physics of the device without the need to take into account the superconducting segment as an effective coupling \cite{cps_flindt_adiabatic,cps_flindt_dynamics} between the QDs. We elucidate the decomposition of the net TE signal into CAR and ECT components, outline the effects of quantum broadening and hybridization, and establish the presence of nonlocal quantum correlations arising at the CPS resonance via the formalism of quantum discord. Thereby, this work provides detailed insights into the gate voltage control of the nonlocal quantum correlations in superconducting-hybrid Cooper pair splitters, revealing new avenues for harnessing quantum correlations in solid-state systems.
\begin{figure}[t]
    \centering
    \includegraphics[height=10cm, width = 0.5\textwidth]{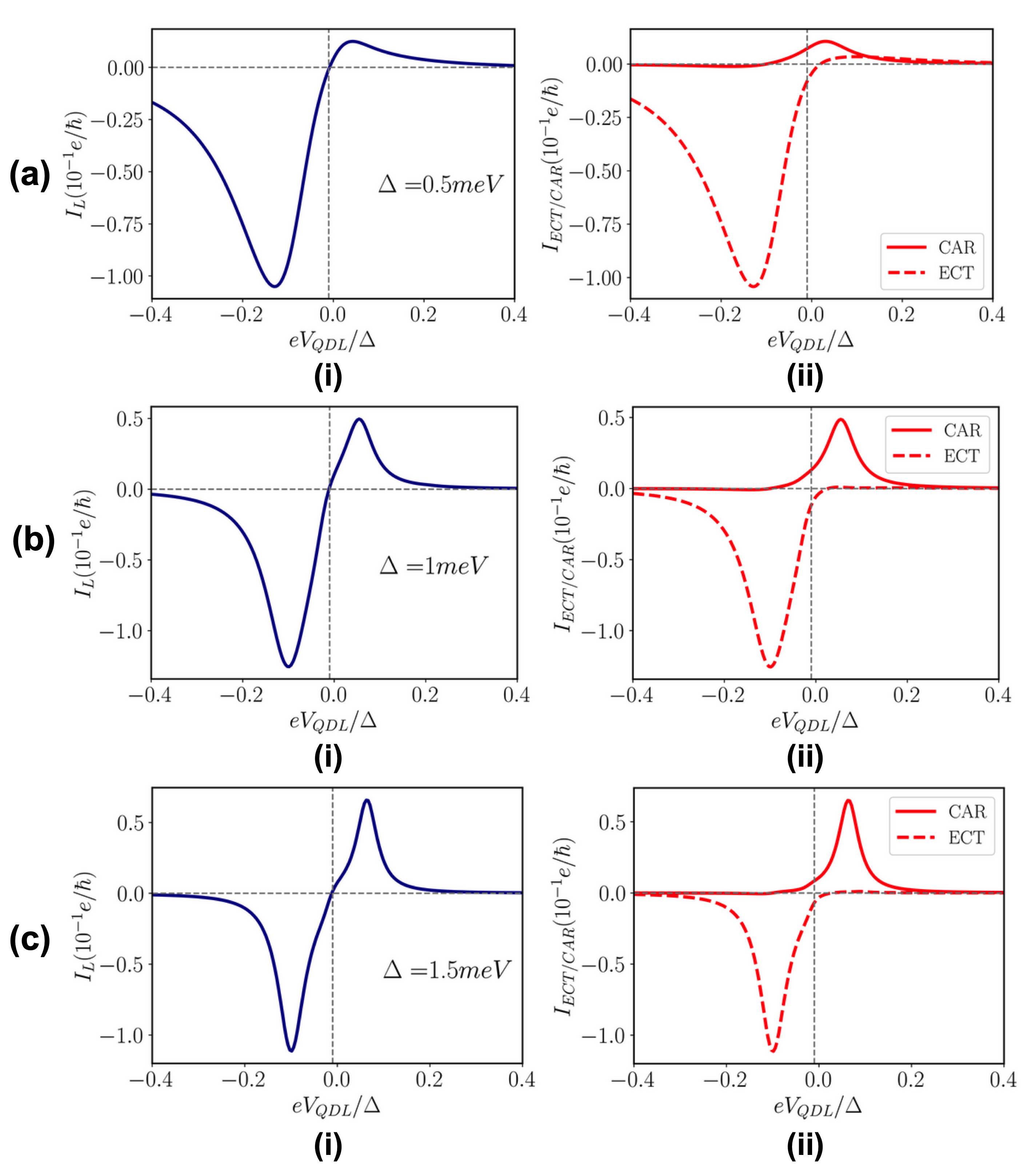}
    \caption{{\bf{Deconstructing the bi-lobe structure of the TE currents.}} Current profiles across the device structure for (a) $\Delta = 0.5$meV, (b) $\Delta = 1$meV and (c) $\Delta = 1.5$meV. (i) Total current through the left contact ($I_L$). (ii) The CAR ($I_{CAR}$) and ECT ($I_{ECT}$) resolved currents. All the curves involve varying the left dot voltage $V_{QDL}$ with a fixed right dot voltage of $eV_{QDR}= -0.1 \Delta$ and a thermal bias of $k_B\delta T = 0.5\Delta$. The chemical potential $\mu = 25$meV, and couplings $t^{QN} = t^{QS} = 0.02t_0$ are held constant as $\Delta$ is varied.}
    \label{fig:fig2}
\end{figure}
\section{Results}

\subsection{Device Setup and Qualitative Reproducibility}

A schematic of the setup is shown in Fig. \ref{fig:fig1}(b,c). The channel region consists of a one-dimensional superconductor sandwiched between two quantum dots, QDL and QDR. This channel region is tunnel coupled to two normal contacts (N) on the left and right sides. The overall setup Hamiltonian is $H =H_N +H_{Ch}+ \sum_{\alpha}H_{QD \alpha-N}$, where $H_{Ch}, H_{N}$, and $H_{QD \alpha-N}$ (with $\alpha = L,R$),  represent the channel Hamiltonian, the Hamiltonian of N contacts, and the coupling Hamiltonians between the N region and QDL(R), respectively. \\
\indent The Hamiltonian of the channel reads, $H_{Ch} = H_{S} + H_{QDL} + H_{QDR} + H_{QDL,S} + H_{QDR,S}$, where $H_S$ is the effective 1D SC Hamiltonian
\begin{eqnarray}
    \begin{aligned}
        H_{S} &=  \sum_{i,\sigma=\uparrow \downarrow}(2t_0 - \mu_0) c_{i\sigma}^\dagger c_{i\sigma}  + \sum_{i,\sigma=\uparrow \downarrow} (t_0 c_{i+1\sigma}^\dagger c_{i\sigma} +h.c.)  \\
        & + \sum_{i}(\Delta c_{i\uparrow}^\dagger c_{i\downarrow}^\dagger + h.c.),
    \end{aligned}
   \end{eqnarray}
where $t_0$ is the hopping parameter within the tight binding model of the 1D-superconductor (see Supplementary), $\Delta$ is the $s$-wave superconducting pairing potential, $\mu_0$ is the electrochemical potential and $h.c.$ is the hermitian conjugate. With a lattice discretization of $a=5$nm, we set the length of the superconductor to $L_s = 100$nm by including 20 sites in the calculation. The hopping potential is then calculated as $t_0 = {\hbar^2}/{(2m^*a^2)}$ evaluating to $t_0 = 184$meV. The chemical potential is set to $\mu = 25$meV unless otherwise stated. The summation runs over the site (spin) index $i (\sigma)$. The other components of the channel Hamiltonian, i.e.,  $H_{QDL} + H_{QDR} + H_{QDL,S} + H_{QDR,S}$ represent the Hamiltonians of the QDL, QDR, the coupling between the QDL, QDR with the SC region. These quantities are defined in detail in the Supplementary. \\
\indent We denote the energy levels of the left (right) QD by $\{\epsilon_{L(R)}\}$, whose positions are varied via the potentials applied at the local gate electrodes. In accordance with the experiment \cite{Splitter_Hakonen}, we neglect the Coulomb repulsion, given that there is no clear experimental evidence of Coulomb blockade in the QDs. This is due to the large overlap of the Cooper pair injector with the QDs which results in a tunnel coupling interface with large capacitance. A gate voltage $V_{QDL(R)}$ is applied at the left (right) QD to shift the local energy levels, $\{\epsilon_{L(R)} - eV_{QDL(R)}\}$, where $e$ is the magnitude of the electronic charge. The QDs are coupled with the nearest neighbour SC sites via spin-conserving coupling strengths $t^{QS}_{L(R)}$, which accounts for the hybridization. \\
\indent For the N contacts, we consider two semi-infinite normal metallic leads in the site basis (See Supplementary) which are coupled with strengths $t^{QN}_{L(R)}$ to the left and right QDs respectively. The semi-infinite structure of the leads is accounted for by the self-energies $\Sigma_{L(R)}$ in the NEGF formalism \cite{Kejri_1} (see Sect.~\ref{METH} and Supplementary). We represent all relevant quantities in the site (spin) Nambu representation $ \hat{\psi}_i^{\dagger}=  \begin{pmatrix} c_{i \uparrow}^{\dagger} & c_{i \downarrow} \end{pmatrix}$, where elements of any matrix at a given site are in the $2 \times 2$ electron-hole Nambu space.\\
\indent We start by noting the experimental trace of the local TE current depicted in Fig.~\ref{fig:fig1}(d)(i), which comprises a series of double lobes, and that each double-lobe corresponds to a crossing pair of levels between QDL and QDR. Hereon, we focus on the single TE lobe, and present results pertinent to the essential physics of the CPS operation, rather than a full simulation of the experiment. For this, we include only a single level on both QDs. The right QD level is kept fixed at resonance with the chemical potential, similar to the setup in Fig. \ref{fig:fig1}(c). 
\indent  Figure \ref{fig:fig1}(d)(ii) shows the theoretically calculated left lead current, which depicts a single sawtooth behavior and a temperature variation consistent with the experimental curves. To identify and verify CAR-dominated parameter regimes, we resort to an information-theoretic quantity exemplifying non-classical correlations known as the quantum discord \cite{quantumdiscord}. Due to the inherent difficulty of computing entanglement in general mixed states, we utilize quantum discord as a proxy for entanglement generation and apply the recently developed method for calculating discord in fermionic systems \cite{fermionicdiscord}. Informally, the discord is defined as the difference between the total and the purely classical correlations, i.e., non-zero discord implies the quantum nature of correlations (see Sect.~\ref{METH} and Supplementary Material). We calculate the quantum discord between the opposite spin modes on the left and right dots, as shown in Fig. \ref{fig:fig1}(d)(iii). \\
\indent Unlike the TE current signal, the discord peaks only at gate voltages coinciding with one of the lobes. As we shall see in more detail, this lobe coinciding with the peak in the discord actually represents CAR contributions to the total current, whereas the ECT dominates the current in the other lobe. The discord thus confirms the presence of a spin-singlet correlation between the two dots in the CAR dominated regions and also showcases the absence of such correlations in ECT dominated regimes. \\
\indent However, it must be noted that while there could be non-zero mutual information between the same spin $\sigma-\sigma$ orbitals of the QDL-QDR, it does not represent any particle entanglement, as it corresponds to correlations of the type $\langle c^{\dagger}_{QDL}c_{QDR}\rangle$. Indeed, we have verified that there is no quantum discord between the $\sigma-\sigma$ orbitals throughout the $V_{QDL}$ axis with our calculations (not shown). On the other hand, the discord computed here corresponds to the correlations of the form $\langle c^{\dagger}_{QDL}c^{\dagger}_{QDR}\rangle$ and is thus treated as a proxy for steady state entanglement generation via the CPS mechanism.
\subsection{Resolving CAR and ECT components}
To analyze the effects of various parameters on the Cooper pair splitting process, it is necessary to first resolve the TE current signal into its CAR and ECT components. In Fig. \ref{fig:fig2}(a)(i), we plot the TE current for a setup with the SC gap $\Delta = 1$meV, a fixed right-dot gate voltage $eV_{QDR} = -0.1\Delta$, thermal bias $k_BT = 0.5\Delta$ and couplings $t^{QN} = t^{QS} = 0.02t_0$. In Fig. \ref{fig:fig2}(a)(ii), We decompose the total TE current into its CAR (solid) and ECT (dashed) components.  \\
\indent These contributions are discernible through the energetics of the underlying subgap transport mechanism. Moving right-to-left along the $V_{QDL}$ axis, when $eV_{QDL}\sim-0.1\Delta$, the left dot level is effectively at $\epsilon_L-eV_{QDL}=0.1\Delta$ and resonates with the right dot level positioned at $\epsilon_R - eV_{QDR}=0.1\Delta$, thereby leading to a maximum of the ECT process which is essentially just a direct tunneling process mediated through the suppressed subgap DOS within the SC region. Next, when $eV_{QDL}\sim0.1\Delta$ the left dot level is effectively at $\epsilon_L-eV_{QDL}=-0.1\Delta$ and satisfies $\epsilon_L-eV_{QDL} = -(\epsilon_R-eV_{QDR}) = 0.1$, at which point the CPS process is assisted by a ready availability of energy levels at $E$ and $-E$, leading to a peak in the CAR process. \\
\begin{figure*}[t]
    \centering
    \includegraphics[height=14cm, width = 0.75\textwidth]{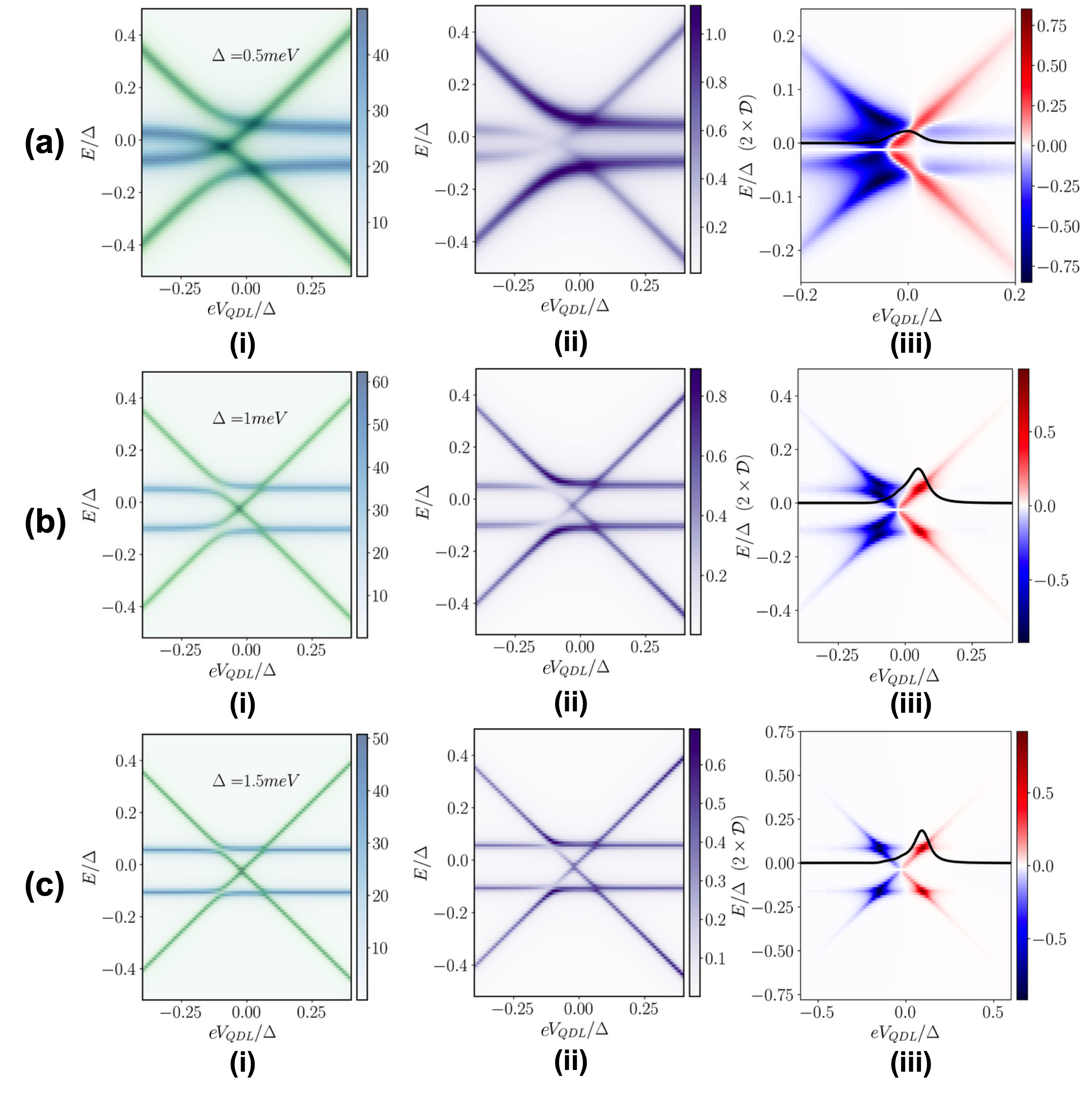}
    \caption{{\bf{Avoided crossings: local density of states and transport spectra.}} For (a) $\Delta = 0.5$meV, (b) $\Delta = 1$meV and (c) $\Delta = 1.5$meV. (i) Superimposed LDOS of QDL (green) and QDR (blue), with the scalebar representing the same maximum intensity. (ii) The DOS of the mediating SC segment. (iii) Energy resolved current spectrum, with the inset curve showing twice the quantum discord $\mathcal{D}$. Parameters are the same as that of Fig.\ref{fig:fig2}. All DOS plots are in the units of $\frac{1}{\Delta}$ and current spectra in units of $\frac{e}{\hbar \Delta}$.}
    \label{fig:fig3}
\end{figure*}
 \begin{figure*}[t]
    \centering
    \includegraphics[height=9cm, width = \textwidth]{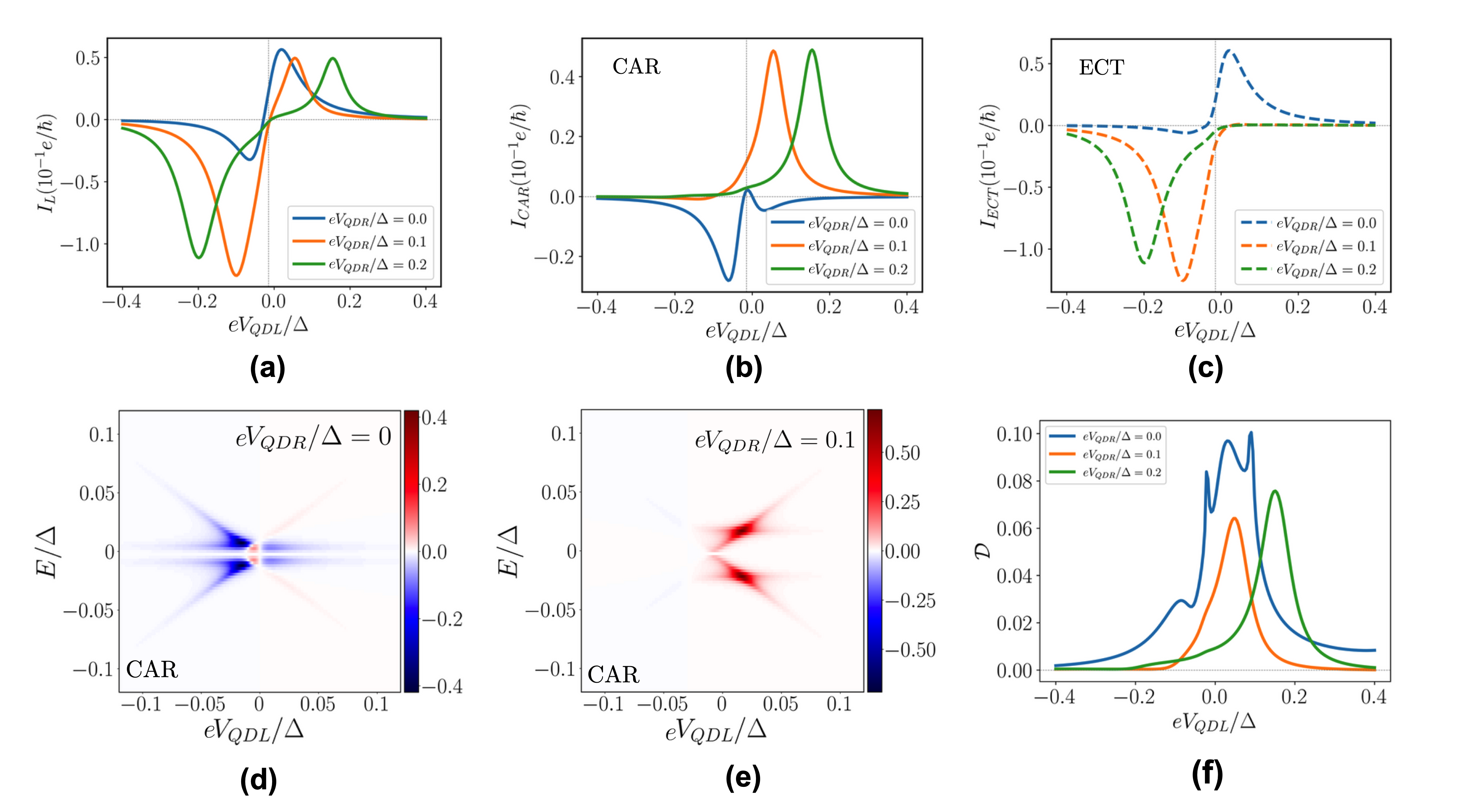}
    \caption{{\bf{Detuning the right quantum dot.}} (a) TE current through the left lead $I_L$ in units of $0.1e/\hbar$. (b-c) Resolved CAR and ECT currents. (d-e) Energy resolved CAR current for $eV_{QDR} = 0
    $ and $eV_{QDR} = 0.1\Delta$ respectively. (f) Quantum discord for $V_{QDR}$ values corresponding to (a). A comparison between the CAR current spectrum in (d)  and (e) clearly shows the flipping of the CAR and ECT lobes with respect to $V_{QDL}$ when $V_{QDR}= 0
    $. This is a non-trivial aspect of level broadening due to hybridization of the QD level.}
    \label{fig:fig4}
\end{figure*}
\indent Including the finite SC segment as an integral part of the analysis, rather than adding effective tunnel coupling between the QDs, as done in master equation approaches \cite{cps_flindt_adiabatic}, brings to fore many nuances. To further elucidate this, in Fig. \ref{fig:fig2}(a-c), we plot the TE current at the left contact for $\Delta = 0.5, 1.0,$ and $1.5$meV respectively. Looking at Fig.~\ref{fig:fig1}(b), for a given placement of $V_{QDR}$, one would expect that the ECT resonances will occur at $E=eV_{QDR}$ and the CAR resonance at $E=-eV_{QDR}$. This is because the ECT resonance occurs because of the spin-conserving tuning process, which takes place when the up- (down-) spin electron (hole) levels align. The CAR resonance occurs at just the opposite energetic condition. We thus note two important subtleties: one, the resonance conditions for the CAR and ECT lobes are \textit{not} exact as seen by the position of the peaks in Fig. \ref{fig:fig2}(a-c). Secondly, Figs. \ref{fig:fig2}(a)(i), (b)(i), and (c)(i) show that as we increase the order parameter, the CAR component increases, with the ECT component remaining nearly unchanged. Moreover, upon increasing $\Delta$, we note that the CAR peak moves closer to the expected value of $eV_{QDR}$ but the location of the ECT peak remains largely unaltered. \\
\indent To gain a qualitative difference between the CAR and ECT peaks, albeit being subgap processes that are suppressed exponentially as the SC length exceeds the coherence length \cite{gen66}, can be understood from the following insight. 
We recall that the coherence length is defined as $\xi_S = {\hbar \nu_f}/{(\pi \Delta)}$. Noting that the coherence length is inversely proportional to the superconducting order parameter, we can expect that CAR processes become more prominent as $\Delta$ increases, which is consistent with many previous studies \cite{Falci_2001,Flatte}. 
To demonstrate this, we vary the coherence length by changing $\Delta$ and keeping the length of the superconducting segment ($L_s = 0.1 \mu$m) constant. In Figs. \ref{fig:fig2}(a)(i), (b)(i) and (c)(i),  we set $\xi_S/L_s = 16.3, 8.1 \text{ and } 5.4$ respectively. We see that, although there is only a nuanced effect on the ECT peaks, the CAR peaks improve dramatically as the coherence length becomes comparable to $L_s$. A naive intuition would be that decreasing $L_s$ would always increase the nonlocal signal; instead, one rather requires $\xi_S \simeq L_s$ to allow for optimal splitting of the CAR and ECT peaks. Moreover, this effect further elucidates the fundamental difference between the CAR and ECT peaks---while both are indicative of SC-mediated transport between the two quantum dots, the CAR is intrinsically linked to the order parameter $\Delta$, whereas the ECT is related to the quasiparticle tunneling processes that conserve the spin. In the viewpoint of the two spatially separated quantum dots, the ECT appears as a spin-conserving hopping process mediated by the superconducting segment.  Understanding the aforementioned subtleties requires a deeper exploration of level broadening and hybridization, which we shall now discuss.
\subsection{Current spectral decomposition}
In the analysis to follow, for our choice of $V_{QDR}=-0.1 \Delta$, the ECT (CAR) resonances occur to the left (right) abscissa axis. 
We observe in Fig.~\ref{fig:fig2} that the ECT curves are asymmetrically distributed about their resonance, whereas the CAR plots are symmetric. Furthermore, we note that as the SC is strengthened from Fig.~\ref{fig:fig2}(a)(ii) to Fig.~\ref{fig:fig2}(c)(ii), the asymmetry in the ECT curves is less pronounced. This, we will show, is, in fact, due to the avoided crossings of the ECT process as $\Delta$ is varied.\\
\indent We illustrate this in Fig. \ref{fig:fig3}, where we show  (i) the local DOS (LDOS) at the QDs, (ii) the total DOS inside the SC region and (iii) the energy resolved current with discord $\mathcal{D}$ superimposed. All DOS plots are in the units of $\frac{1}{\Delta}$ and current spectra in units of $\frac{e}{\hbar \Delta}$. The results are shown for the same set of superconducting strengths as in Fig. \ref{fig:fig2}. As we move from (a) to (c), the $\Delta$ increases, the influence of the QDs on the SC segment is minimal, which manifests as a sharper DOS with decreased broadening. \\
\indent The QDL LDOS (green, cross-like) is superposed with the QDR LDOS (blue, horizontal line-like) in Fig. \ref{fig:fig3}(a-c)(i), which shows avoided crossings at both the ECT and CAR points. The avoided crossing represents the states of the QD levels in the presence of the SC, with a larger level repulsion when $\Delta$ becomes weaker, indicating a considerable mixing of SC and QD states. This typically happens when the coherence length is large (smaller pairing). Also, such avoided crossings are a generic signature of CAR and ECT processes in a QD-SC-QD setup \cite{cps_flindt_adiabatic} due to two-level atom like transitions between the empty and the singlet states of the two QDs (see Supplementary for further discussion). While the CAR crossing behaves exactly as pointed out in an earlier work \cite{cps_flindt_adiabatic}, the ECT crossing shows non-trivial behavior as the $\Delta$ becomes stronger, as a direct consequence of contact induced level broadening. \\ 
\indent First, we make note that the CAR anticrossings are far less discernible than the ones related to the ECT processes. When we peek into the SC-DOS shown in Fig.~\ref{fig:fig3}(iii) column, for the CAR process, we note an equal increase in the QDL and the QDR branches of the SC-DOS. With increasing $\Delta$, the SC segment becomes less influenced by the QD and the CAR magnitude increases with the resonant peak becoming stronger, as noted previously in Fig. \ref{fig:fig2}. This can be clearly seen with the increased contrast of the CAR spectrum as $\Delta$ increases. This effect is also illustrated by the sharpening of the discord curve as we move from Fig.~\ref{fig:fig3}(a) to (c), further highlighting the {strength} of information-theoretic quantities like discord. By looking at the energy-resolved conductance spectra of the $\Delta = 0.5$meV and $\Delta = 1$meV cases in Fig.~\ref{fig:fig3}(a-b)(iii), it is perhaps difficult to ascertain the nonlocality of the signal obtained. But, the discord computed in the two cases clearly outlines a unique bias point of maximum entanglement generation in Fig.~\ref{fig:fig3}(b)(iii). Intuitively, this is owing to the fact that the discord is a highly non-linear function of the density matrix, while the conductance spectra is linear in the density matrix. This suggests that, although the net CAR and ECT signals are accessible, they do not uniquely reveal the QDL voltage point at which the two dots are maximally correlated. In contrast, analyzing the discord offers a clearer insight into this effect. We further clarify this in the next section.\\
\indent As mentioned earlier, the CAR current magnitude increases \cite{Falci_2001}, while the ECT remains largely the same. 
While the thermoelectric bias depends on $\Delta$, this only affects the currents weakly. Looking carefully at the CAR resonances for the current spectrum along the column Fig.\ref{fig:fig3}(iii), we notice specifically from Fig.\ref{fig:fig3}(a)(iii) that there is a current back flow contribution as noted from the blue striped region. This clearly corresponds to processes that involve mixing of the QD levels with those of the SC, thereby promoting normal reflections in addition to the Andreev processes. This leads to the weakening the CAR leading to a smaller TE current. The CAR processes get stronger with sharper spectral resolution as the $\Delta$ is increased, leading to an increase in the total current, which is an integral over the energy axis at a given $V_{QDL}$.
\subsection{Detuning the QDR}
Moving further, we analyze the effects of detuning the right dot with respect to the electrochemical potential of the system. Figure \ref{fig:fig4}(a) demonstrates the TE current through the left lead $I_L$ as a function of left dot gate voltage $V_{QDL}$ for $eV_{QDR} \in \{0.0 , 0.1, 0.2\} \Delta$. The characteristic saw-tooth widens upon increasing $V_{QDR}$. Upon resolving the left lead current in Fig. \ref{fig:fig4}(b) and (c), it is clear that the CAR and ECT peaks of the current move away from the resonance point of the hybridised dot level. Upon further inspection of the resolved CAR and ECT currents, we note that for the case of zero detuning, the peaks of CAR and ECT are \textit{interchanged}, accompanied by a parity crossing in the currents. \\
\indent The parity crossing observed in the currents is characteristic to a TE bias when only a temperature gradient applied. The electron and hole currents flip when the chemical potential falls precisely at the mid-gap of a given spectrum. In our case, due to the finite hybridization between either dot and the SC region, this does not happen at the apparent mid-gap symmetry point, which is $eV_{QDR}=0$. This can be further elucidated in Fig. \ref{fig:fig4}(d-e), where we plot the energy resolved spectrum of the CAR currents for (d) $eV_{QDR}/\Delta=0$ and (e) $eV_{QDR}/\Delta=0.1$. \\
\indent In Fig. \ref{fig:fig4}(d), we note two parity changes upon moving from left to right along the $V_{QDL}$ axis and a positive CAR peak on the left side, which moves over to the right side with a single parity crossing upon increasing the detuning to $0.1\Delta$ as in Fig. \ref{fig:fig4}(e). Thus, there exists a critical detuning magnitude that is dependent on the SC region and the coupling strengths, after which, the expected behavior sets in for the CAR and ECT peaks. This interchange of the CAR and ECT lobes is crucial to understand when interpreting the experimental current traces. \\
\indent Lastly, Fig. \ref{fig:fig4}(f) shows the discord for corresponding $V_{QDR}$ values. While the discord peak moves with the CAR maxima, a critical difference is noted between the $eV_{QDR}/\Delta=0$ and the $eV_{QDR}/\Delta \in \{0.1,0.2\}$ cases, with the location of the side-peak moving from the left to the right respectively, thereby emphasizing the physics of hybridization due to the contact broadening. We further observe that, for the case of \( V_{QDR} = 0 \), the maxima of the CAR and discord curves do \textit{not} coincide along the \( V_{QDL} \) axis. This underscores the importance of analyzing discord along with transport signals. In the pathological case with reversed CAR and ECT peaks, we find that the bias point of maximal quantum discord does not align with the CAR peak. Instead, as shown in Fig.~\ref{fig:fig4}(b) and (f), the maximum of the discord corresponds to a local minimum in the CAR curve at \( V_{QDL} > 0 \). As we move forward, the expected alignment between the point of maximal entanglement and the CAR peak resumes for \(eV_{QDR}/\Delta \gtrsim 0.1\).\\ 
\indent Thus, a combined study of dc-transport signatures and operationally functional information-theoretic quantities, such as quantum discord, enables a comprehensive understanding of the TE sawtooth. This approach provides clear insight into both direct tunneling processes and CPS entanglement generation.
\section{Discussions and Inferences} We conducted a detailed analysis of the observed local TE currents \cite{Splitter_Hakonen}, which revealed crucial insights into the operating regimes and the complex nature of the correlations. The behaviors of the CAR and ECT anti-crossings at different gate voltages highlighted the significant roles of hybridization with the SC segment and level broadening. A detailed current spectral analysis connected these features to the observed transport signals, examining parity reversal and the shifted resonances of the CAR processes. Finally, we established the presence of nonlocal correlations in the CAR current signal through the quantum discord formalism. By interpreting discord as a proxy for entanglement generation, we gained novel insights into the CPS process that could not be obtained through transport signals alone. We believe that the experimental deduction of quantum discord would be an integral next step in determining the nonlocal correlations of the CPS process, and the groundwork for witnessing quantum discord \cite{RMP_Dis,Lowe}, although in a nascent stage, are already being developed.  \\
\indent Another interesting future direction involves integrating the full spectral information available through our calculations into a time-dependent quantum master equation \cite{futurework_integratenegf} to improve recently proposed driving protocols to maximize CPS efficiency \cite{cps_flindt_adiabatic,cps_flindt_dynamics}. Furthermore, more precise calculations of the entanglement generation in CPS devices await a quantum trajectory unraveling of the dynamics \cite{daley2014quantum}. Realizing CPS in two-dimensional topological materials is another promising direction \cite{CPS_QSH,Midha_2024}. Additionally, it remains to be seen whether the mechanism of spin-valley locking in certain systems, for example, bilayer graphene or transition-metal dichalcogenide quantum dots, might be used to engineer more precise control over the splitting process in such devices, as well as harness nonlocal valley entanglement.

\section{Methods} \label{METH}
We employ the Keldysh-NEGF formalism for the transport calculations. Broadly, it involves calculation of the retarded Green's function from which the other quantities of interest are derived. The detailed procedures are addressed in the Supplementary Material as well as many recent works \cite{Kejri_1}. In the sub-gap transport regime, one evaluates the transmissions for the ECT and the CAR processes, as given by
\begin{equation}
\begin{aligned}
&T^{e(h)}_{ECT}(E) = {\rm Tr }\left(\Gamma_{L}^{e e(h h)} G^{r} \Gamma_{R}^{e e(h h)} G^{a}\right) \\
&T^{e(h)}_{CAR}(E) = {\rm Tr }\left(\Gamma_{L}^{e e(h h)} G^{r} \Gamma_{R}^{h h(e e)} G^{a}\right),
\end{aligned}
\end{equation}
where, $G^r~(G^a)$ is the retarded (advanced) Green's function, $\Gamma_{L(R)}$ is the broadening matrix associated with the left (right) contact with the superscripts $ee(hh)$ denoting the electron(hole) sub-sectors in the Nambu representation. The net spectral current at the contacts is a sum of these transmissions weighed appropriately with the Fermi-Dirac distribution differences at the contacts \cite{Kejri_1}, \\
\begin{equation}
\begin{aligned}
    & I(E) = I^e(E) - I^h(E),  \\
    & I^{e(h)}(E) = T^{e(h)}_{CAR}(E)[f(E \mp eV_L, T_L) - f(E\pm eV_R, T_R)] \\
    & + T^{e(h)}_{ECT}(E)[f(E \mp eV_L, T_L) - f(E \mp eV_R, T_R)] 
\end{aligned}
\end{equation}
In our study, we apply just a thermal-bias, caused by a temperature difference $\Delta T = T_L - T_R$, and therefore we set $V_L = V_R = 0$, which is precisely why we drop contributions of the Local Andreev processes since they are proportional to the Fermi function difference on the same contact which is exactly zero in the absence of voltage bias. The total current is obtained by integration of the spectral current $I(E)$ over energy. The exact numerical values of NEGF related quantities, such as the number of sites, chemical potential, imaginary damping parameter etc., are provided in detail in our Supplementary Material.\\
\indent Additionally, to characterize the nonlocal quantum correlations between spatially separated left and right QDs, we employ the formalism of \textit{quantum discord} \cite{quantumdiscord}. This quantity is defined as
\begin{equation}
    \mathcal{D}(A:B) := S(A:B) - I(A:B),
\end{equation}
where $S(A:B)$ is the quantum mutual information between A and B encoding all possible correlations, and $I(A:B)$ is the \textit{maximum} classical mutual information, maximized over all possible measurements on $B$. Computing this quantity typically requires an optimization over possible measurements to calculate the maximum classical correlations. However, owing to the fermionic parity-superselection rule, this calculation greatly simplifies \cite{fermionicdiscord} (see Supplementary Material). The NEGF equations \cite{Kejri_1,Arora_2024,Midha_2024} are first solved to get the steady-state behavior of the system at individual energies $E$, and then a steady-state correlation matrix is computed by integrating electron density lesser Green's function $G^n(E)  = -i G^{<}(E)$ (see the Supplementary for details) along the energy axis, such that  
\begin{equation}
    -i \langle \hat{\psi}_i^{\dagger}  \hat{\psi}_j \rangle = \frac{1}{2\pi}\int dE [G^n(E)]_{i,j}.
\end{equation} 
This is a correlator matrix in the Nambu representation with the off-diagonal terms representing the superconducting pairing amplitude. 
Subsequently, the two-mode occupation-basis density matrix for the up-spin mode of the left QD and the down-spin mode of the right QD is computed using the above steady-state correlation matrix while accounting for correlations within the gap (see Supplemetary Material).\\
\indent Quantum discord, defined as the difference between the quantum and maximal classical mutual information, is an indicator of quantum correlations between two systems. Following \cite{fermionicdiscord}, we compute the two-orbital fermionic discord between the up-spin orbital of the QDL and the down-spin orbital of the QDR to probe correlations in the CPS (see Supplementary Material).\\

\section*{Acknowledgements}
We thank D. Golubev, M. Kumar, Z. Tan, and V. Vinokur for useful discussions and correspondence. This work was supported by the Research Council of Finland (RCF) Project Nos. 341913 (EFT), 352926 (CoE, Quantum Technology Finland). The author BM wishes to acknowledge the support by the Science and Engineering Research Board (SERB), Government of India, Grant No. MTR/2021/000388. The authors SM and BM acknowledge support of the Dhananjay Joshi Foundation from an Endowment to IIT Bombay. The author BM and PH acknowledge funding from the InstituteQ and Aalto University visiting program and the MEC Global funding of Aalto University. AAZ acknowledges support from the QuantERA II Programme that has received funding from the European Union’s Horizon 2020 research and innovation programme under Grant Agreement No 101017733. 

\section*{Data Availability}
The data that support the plots within this paper and other findings of this study are available from the corresponding author upon reasonable request.

\section*{Code Availability}
The codes generated during the simulation study are available from the corresponding author upon reasonable request.


\section*{Author Contributions}
AA and SM developed the theoretical framework of the device, performed the numerical simulations, and devised the connection to quantum information. The ideas germinated on the basis of experimental works done in the Aalto University group and based on extensive discussions between BM, PH and AZ. All authors contributed to the analysis of the results and writing of the paper.

\section*{Competing Interests}
The authors declare that there are no competing interests.

\bibliography{Bib_RS} 
\onecolumngrid
\renewcommand{\figurename}{Supplementary Figure}
\setcounter{figure}{0}    
\setcounter{section}{0}    

\newpage

\section*{Supplementary Information}

  \section{\label{sec:setup} Setup and Formalism}
    A schematic of our setup is shown in Fig. 1(c) of the main text. The device consists of a superconducting (SC) region sandwiched between two quantum dots (QD). The QDs are tunnel coupled to two normal contacts (N) on the left and right. The length of the SC region is smaller than the SC coherence length to ensure that non-local sub-gap processes like ECT and CAR can manifest with considerable magnitude. Each QD may have multiple levels with level positions and separations independent of each other. We denote the levels of the left(right) QD by $\{\epsilon_{L(R)}\}$. A gate voltage $V_{QDL(R)}$ can be applied to the left(right) QD to control its local chemical potential and shift the levels to $\{\epsilon_{L(R)} - eV_{QDL(R)}\}$, where $e$ is the electronic charge. The Hamiltonian of the QD-SC-QD region is given by 
    \begin{equation}
    \begin{aligned}
        H &= H_{S} + H_{QDL} + H_{QDR} + H_{QDL-S} + H_{QDR-S}
    \end{aligned}
    \end{equation}
    where, $H_{S}$ is the Hamiltonian of the SC region
    \begin{equation}
        H_{S} = \sum_{i=1}^{N_s} \sum_{\sigma=\uparrow \downarrow}(2t_S - \mu_S) c_{i\sigma}^\dagger c_{i\sigma}  - \sum_{i=1}^{N_s -1} \sum_{\sigma=\uparrow \downarrow}t_0c_{i+1\sigma}^\dagger c_{i\sigma}
         + \sum_{i=1}^{N_s} \Delta c_{i\uparrow}^\dagger c_{i\downarrow}^\dagger + h.c.
    \end{equation}
    $N_s$ is the number of sites in the SC region, $c_{i\sigma}^{(\dagger)}$ are the annihilation(creation) operators at site $i$, $\mu_S$ is the chemical potential, $\Delta$ is the SC gap parameter and $t_S = \frac{\hbar^2}{2m^*a^2}$ is the tight-binding hopping parameter with electron reduced mass $m^*$ and lattice spacing $a$. The Hamiltonians of the left(right) dot read,
    \begin{equation}
        H_{QDL(R)} = \sum_{i=1, \sigma=\uparrow \downarrow}^{N_{dL(R)}} (\epsilon_{L(R)} - eV_{QDL(R)})d_{i\sigma L(R)}^\dagger d_{i\sigma L(R)}
    \end{equation}
    $N_{dL(R)}$ are the number of levels in the left(right) dot, operators $d_{i\sigma}^{(\dagger)}$ annihilate(create) electrons in a state $i$ with spin $\sigma$ in the respective dot and $\epsilon_{L(R)}$ and $eV_{QDL(R)}$ define the energy levels of the dot. \\
    \indent Further, we have $H_{QDL(R) - S}$, which is the Hamiltonian that couples either dot to theSC region,
    \begin{equation}
        H_{QDL(R)-S} = \sum_{i=1, \sigma=\uparrow \downarrow}^{N_{dL(R)}} (t^{QS}_{L(R)} d_{i\sigma L(R)}^\dagger c_{1(N_S)\sigma} + h.c.)
    \end{equation}
    here, $t^{QS}_{L(R)}$ denotes the hopping between the left(right) QD and the SC. For the contacts, we consider two semi-infinite metallic leads in the eigenbasis with the Hamiltonian defined as
    \begin{equation}
    \begin{aligned}
        H_{N} &= \sum_{i, \sigma=\uparrow \downarrow} (2t_N - \mu_N) b_{i\sigma}^\dagger b_{i\sigma}  - \sum_{i, \sigma=\uparrow \downarrow}t_Nb_{i+1\sigma}^\dagger b_{i\sigma} + h.c.
    \end{aligned}
    \end{equation}
    where, $b_{i\sigma}^{(\dagger)}$ are the annihilation(creation) operators in the contacts, $\mu_N$ is the chemical potential of the leads and $t_N$ is the tight-binding hopping parameter. The coupling between the semi-infinite leads and the QD are mediated by a Hamiltonian of the form 
    \begin{equation}
        H_{QDL(R)-N} = \sum_{i=1, \sigma=\uparrow \downarrow}^{N_{dL(R)}} (t^{QN}_{L(R)} d_{i\sigma L(R)}^\dagger b_{1\sigma} + h.c.)
    \end{equation}
    here, $t^{QN}_{L(R)}$ denotes the hopping between the left (right) QD and the the left (right) contact, which is in-turn taken into account via the contact self-energies which are calculated iteratively using the surface Green's function method.

\begin{table}[h]
\centering
\begin{tabular}{|c|c|c|}
\hline
Parameter & Symbol & Value(s) \\ \hline
Discretization & $a$ & $5$nm \\ \hline
Superconductor length & $L_s$ & $100$nm \\ \hline
No. of sites in SC & $N = \frac{L_s}{a}$ & 20 \\ \hline
Superconducting order & $\Delta$ & $\{0.5,1.0,1.5\}$meV \\ \hline
Chemical potential & $\mu$ & $25$meV\\ \hline
Hopping parameter & $t_0$ & $184$meV \\ \hline
QD-N coupling & $t^{QN}$ & $0.02t_0$ \\ \hline
QD-S coupling & $t^{QS}$ &  $0.02t_0$ \\ \hline
Effective mass & $m^*$ &  $0.023m_e$ \\ \hline
Infinitesimal damping parameter & $\eta$ & $1 \times 10^{-12}$ \\ \hline
\end{tabular}
\caption{Summary of the parameters}
\end{table}
From the table above, we note that the choice of the number of SC sites is even. The odd-even effects are usually an issue in any tight-binding model. In the case of superconducting segments, one typically uses an even number of superconducting sites to avoid the oscillatory behavior of the results \cite{Odd_Even} in the limit of a small number of sites. Furthermore, the location of resonances of subgap states is affected due to the momentum matching conditions that arise due to discretization. This changes the location of the CAR and ECT peaks and may modify the results for small $N$. 

\section{Quantum Transport through the QD-S-QD system}
    For transport calculations we use the Keldysh non-equilibrium Green's functions (NEGF) formalism, described in detail herein. Under the NEGF method, the semi-infinite leads and their coupling to the device are accounted through self-energies $\Sigma_{L(R)}$ of the left(right) contact. These self-energies are calculated recursively from the surface Green's function of the leads -- a standard approach in NEGF calculations. We refer the readers to the supplementary material of \cite{Arora_2024, Kejri_1, duse2021, sriram2019}. Given the Hamiltonian of the channel $H_{Ch}$ and the self-energies of the metallic contacts on the left(right) $\Sigma_{L(R)}$ in their matrix representation, the retarded Green's function matrix at energy $E$ for the QDL-SC-QDR system is calculated as 
    \begin{equation}
        \begin{aligned}
            G^r(E) = \left((E +i\eta)I - H_{Ch} - \Sigma_L - \Sigma_R \right)^{-1}
        \end{aligned}
    \end{equation}
    where, $\Sigma_{L(R)}$ are the contact self-energies, $I$ is the identity matrix of dimension $2(N+2) \times 2(N+2)$, with $N$ denoting the number of superconducting sites, and $\eta>0$ is a small damping parameter. The advanced Green's function is then defined as $G^a \equiv [G^r]^{\dagger}$. \\
    \indent It is worth noting that, the definition of the above Hamiltonian in the BdG representation is in the composite electron-hole basis. One can also use the ``site basis'' to define the $[\alpha]$ and $[\beta]$ matrices corresponding to the on-site and hopping sections. In the composite basis, any matrix $[A]$ is reorganized in a block form as 
\begin{equation}
[A] = \begin{pmatrix} [A^{ee}] & [A^{eh}] \\ [A^{he}] & [A^{hh}] \end{pmatrix}, 
\label{Block}
\end{equation}
where, each block is now $2(N+2) \times 2(N+2)$ and the superscripts $e$ and $h$ represent the electron and hole indices respectively. All elements of the Green's function, self energies, correlation matrices etc., will have this block form. This form helps us to gain some analytical insights into the calculations, while the actual coding may be done in any convenient representation. \\
\indent  In the sub-gap transport regime, transmission manifests in three distinct processes -- Andreev transmission ($T_A$), elastic co-tunneling ($T_{ECT}$) and crossed-Andreev transmission ($T_{CAR}$), given by
    \begin{equation}
\begin{aligned}
&T_A = {\rm Tr }\left(\Gamma_{L}^{e e(h h)} G^{r} \Gamma_{L}^{h h(e e)} G^{a}\right) \\
&T_{ECT} = {\rm Tr }\left(\Gamma_{L}^{e e(h h)} G^{r} \Gamma_{R}^{e e(h h)} G^{a}\right) \\
&T_{CAR} = {\rm Tr }\left(\Gamma_{L}^{e e(h h)} G^{r} \Gamma_{R}^{h h(e e)} G^{a}\right)
\end{aligned}
\end{equation}
where, $G^a = (G^r)^\dagger$ is the advanced Greens function, $\Gamma_{L(R)} =i[\Sigma_{L(R)} - \Sigma_{L(R)}^{\dagger} ]$ are the broadening matrices and $\Gamma_{L(R)}^{ee(hh)}$ are the electron(hole) sub-sectors of $\Gamma_{L(R)}$. The energy resolved current at the left-lead can be calculated as 
\begin{equation}
    \begin{aligned}
        I_{L}(E) &= T_A(E)[f_L^e(E) - f_L^h(E)]\\
        &+ T_{ECT}(E)[f_L^e(E) - f_R^e(E)]\\
        &+ T_{CAR}(E)[f_L^e(E) - f_R^h(E)]
    \end{aligned}
\end{equation}
where, $f_{L(R)}^{e/h}(E) = f(E\pm eV_{L(R), T_{L(R)}})$ ($\pm$ corresponding to $e/h$) and $f(E, T)$ is the Fermi-Dirac distribution at energy $E$ and at temperature $T$. The net current at the left contact can now be computed as,
\begin{equation}
    I_{L} = \int_{-\infty}^{\infty}I_{L}(E)dE
\end{equation}
The density of states is derived from the spectral function which is defined as
\begin{equation}
    A(E) = \iota[G^r(E) - G^a(E)]
\end{equation}
which gives,
\begin{equation}
    \text{LDOS}(x;E) = \frac{1}{2\pi} A(x,x;E)
\end{equation}
\section{\label{sec:discord}Correlations and QDL-QDR density matrix}
Given the broadening matrices $\Gamma_{L(R)}$, one can further compute the in-scattering functions as,
\begin{equation}
 \Sigma^{in}_{L(R)}(E) = f(E,V_{L(R)},T_{L(R)}) \cdot \Gamma_{L(R)}.
\end{equation}
Coupling this with the previously calculated Green's functions, we get the \textit{correlation matrix} $G^n$ as, 
\begin{equation}
    G^n(E) = G^r\Sigma^{in}G^a
\end{equation}
where $\Sigma^{in} = \Sigma^{in}_L + \Sigma^{in}_R$ is the net in-scattering function. This matrix encodes the quadratic correlators at an energy $E$,
\begin{equation}
    [G^n(E)]_{ij\alpha\beta} \sim \langle c^{(\dagger)}_{i\alpha}c^{(\dagger)}_{j\beta}\rangle(E).
\end{equation}
The aforementioned Green's function encodes the nonequilibrium correlators, which in the original parlance is related to the lesser Green function $G^n(E) = -i G^{<} (E)$, in the steady state of the system perturbed by the contacts. Given this, we compute the correlation matrix by integrating along the energy space as follows,
\begin{equation}\label{eq:corrmat}
    [\langle c^{(\dagger)}_{i\alpha}c^{(\dagger)}_{j\beta}\rangle]_{ij\alpha\beta} = \frac{1}{2\pi}\int [G^n(E)]dE.
\end{equation}
With this, we can obtain the specific correlation matrix in the QDL-QDR subspace. We further proceed by obtaining the two-mode density matrices in the $\{\ket{0},\ket{1}\}\otimes \{\ket{0},\ket{1}\}$ occupation basis for the up (down) -spin mode on the QDL and the down (up)-spin mode on the QDR. We recap the general procedure for calculating the density matrix and subsequently the two-mode discord, as outlined in \cite{fermionicdiscord}. \\
\indent Given the correlation matrix $[\langle c^{(\dagger)}_{i}c^{(\dagger)}_{j}\rangle]_{ij}$ of a two-mode system, we compute the density matrix in the \\ $\left\{c_i^{\dagger} c_j^{\dagger}|0\rangle, c_i^{\dagger}|0\rangle, c_j^{\dagger}|0\rangle,|0\rangle\right\}$ basis as \cite{di2018fermionic},
\begin{equation}
 \rho_{i j}=\left(\begin{array}{cccc}
\langle c_i^{\dagger} c_i c_j^{\dagger} c_j\rangle & 0 & 0 & \langle c_j c_i\rangle \\
0 & \langle c_i^{\dagger} c_i c_j c_j^{\dagger}\rangle & \langle c_j^{\dagger} c_i\rangle & 0 \\
0 & \langle c_i^{\dagger} c_j\rangle & \langle c_i c_i^{\dagger} c_j^{\dagger} c_j\rangle & 0 \\
\langle c_i^{\dagger} c_j^{\dagger}\rangle & 0 & 0 & \langle c_i c_i^{\dagger} c_j c_j^{\dagger}\rangle
\end{array}\right).
\end{equation}
Now, since our theory is non-interacting, we apply Wick's theorem to compute the four-term averages as follows:
\begin{align}
    \langle c_i^{\dagger} c_i c_j^{\dagger} c_j\rangle &=  \langle c_i^{\dagger}  c_i\rangle \langle c_j^{\dagger} c_j\rangle - \langle c_i^{\dagger}  c_j^{\dagger}\rangle \langle c_i c_j \rangle - \langle c_i^{\dagger}  c_j\rangle \langle c_j^{\dagger} c_i\rangle  \\ 
    \langle c_i^{\dagger} c_i c_j c_j^{\dagger}\rangle &= \langle c_i^{\dagger}  c_i\rangle (1-\langle c_j^{\dagger} c_j\rangle) + \langle c_i^{\dagger}  c_j\rangle \langle c_j^{\dagger} c_i\rangle + \langle c_i^{\dagger}  c_j^{\dagger}\rangle \langle c_i c_j \rangle \\ 
    \langle c_i c_i^{\dagger} c_j^{\dagger} c_j\rangle &= (1-\langle c_i^{\dagger}  c_i\rangle) \langle c_j^{\dagger} c_j\rangle + \langle c_i^{\dagger}  c_j\rangle \langle c_j^{\dagger} c_i\rangle + \langle c_i^{\dagger}  c_j^{\dagger}\rangle \langle c_i c_j \rangle \\ 
    \langle c_i c_i^{\dagger} c_j c_j^{\dagger}\rangle &= (1-\langle c_i^{\dagger}  c_i\rangle)(1-\langle c_j^{\dagger} c_j\rangle) - \langle c_i^{\dagger}  c_j^{\dagger}\rangle \langle c_i c_j \rangle - \langle c_i^{\dagger}  c_j\rangle \langle c_j^{\dagger} c_i\rangle
\end{align}
Each of the two-term averages are obtained through suitable integration the correlation Green's function as in Eq.~\ref{eq:corrmat} within the domain $|E| < \Delta$ to account for all the sub-gap correlations, . Moreover, these averages are over the non-equilibrium steady state of the channel, with $\langle A\rangle$ for any two-term operator $A$ denoting $\text{Tr}[A\rho_{\text{NESS}}]$, where, $\rho_{\text{NESS}}$ is the non-equilibrium steady-state density matrix.

The quantum discord is defined as the difference between two measures of information shared between systems $A$ and $B$. Given the maximal classical mutual information $I(A:B)$ and the quantum mutual information $S(A:B)$, we compute the discord $\mathcal{D}(A:B)$ as
\begin{equation}
    \mathcal{D}(A:B) := S(A:B) - I(A:B),
\end{equation}
where the quantum mutual information is given as
\begin{equation}
    S(A:B) = S(A) + S(B) - S(A,B),
\end{equation}
where $S(A/B)$ are the von Neumann entropies of the $A/B$ subsystems, and $S(A,B)$ is the joint entropy of the $AB$ system. Further, the optimized classical mutual information is defined as,
\begin{equation}
    I(A:B) = \max_{\{\Pi_k^B\}}S(A) - S(A,B|\{\Pi_k^B\}),
\end{equation}
where, $\Pi_k^B$ is a measurement onto the B subsystem, which is optimized to learn the maximal information about the A subsystem. Formally, 
$$
S\left(\rho^{(A, B)} \mid\left\{\Pi_k^{B}\right\}\right)=\sum_k p_k S\left(\rho_k^{A B}\right),
$$
where, $$\rho_k^{(A, B)}=\frac{1}{p_k} \Pi_k^{B} \rho^{AB} \Pi_k^{B}$$ is a possible post-measurement state and $$p_k=\operatorname{tr}\left(\Pi_k^{B} \rho^{AB} \Pi_k^{B}\right)$$ is the probability of getting the measurement outcome $k$. Now, it was pointed out that the optimization inherent in the computation of quantum discord for fermionic systems is ruled out because of the parity-superselection rule \cite{fermionicdiscord}. Ignoring the superselection rule and carrying out the optimization can lead to an over-estimation of the correlations. Thus, the only possible measurements are,
$$
\Pi_0=c_j c_j^{\dagger}, \quad \Pi_1=c_j^{\dagger} c_j,
$$
and the only possible post-measurement states are
$$
\begin{aligned}
\rho_0^{(A, B)} & =\frac{1}{\rho_1+\rho_3}\left(\begin{array}{cccc}
\rho_1 & 0 & 0 & 0 \\
0 & 0 & 0 & 0 \\
0 & 0 & \rho_3 & 0 \\
0 & 0 & 0 & 0
\end{array}\right), \\
\rho_1^{(A, B)} & =\frac{1}{\rho_2+\rho_4}\left(\begin{array}{cccc}
0 & 0 & 0 & 0 \\
0 & \rho_2 & 0 & 0 \\
0 & 0 & 0 & 0 \\
0 & 0 & 0 & \rho_4.
\end{array}\right)
\end{aligned}
$$
Because of this, the conditional entropy can be written as
$$
S\left(\rho^{(A, B)} \mid\left\{\Pi_k^{(B)}\right\}\right)=S\left(\mathcal{Z}\left(\rho^{(A, B)}\right)\right)-S\left(\rho^{(B)}\right),
$$
where $\mathcal{Z}(\rho)$ is the dephasing channel, which destroys the off-diagonal elements of $\rho$. Finally, the quantum discord can be written as
$$
\mathcal{D}(A, B)=S\left(\mathcal{Z}\left(\rho^{(A, B)}\right)\right)-S\left(\rho^{(A, B)}\right).
$$
 \section{\label{sec:avoided-cross}CAR and ECT avoided crossings}
We explain the CAR-ECT avoided crossings using an effective double quantum dot (DQD) model. The SC region in a QD-S-QD setup can be thought of as a bridge that mediates interaction between the two QDs. In the case with single-level QDs, the coupling between the dots can be considered as a perturbation to the double quantum dot (DQD) system and can be integrated out to yield an effective Hamiltonian \cite{flindt-dynamic, dvir-eff-hamil}.  
\indent Near the ECT point $\epsilon_L = \epsilon_R$, and the expectation values $\langle \frac{1}{\sqrt{2}}(d_{L\uparrow}^\dagger d_{R\downarrow}^\dagger - d_{L\downarrow}^\dagger d_{R\uparrow} ) \rangle = \langle \frac{1}{\sqrt{2}}(d_{R\downarrow} d_{L\uparrow} - d_{R\uparrow} d_{L\downarrow} ) \rangle = |0\rangle\langle S| = |S \rangle\langle 0| = 0$, thereby restricting transitions between the singlet and the unoccupied state. The effective Hamiltonian can then be written as 
\begin{equation}
    H_{eff} = \sum_{\sigma = \uparrow \downarrow} \epsilon_L d_{L\sigma}^\dagger d_{L\sigma} + \epsilon_R d_{R\sigma}^\dagger d_{R\sigma} +  \Gamma_{ECT} d_{L\sigma}^\dagger d_{R\sigma} + h.c.,
\end{equation}
which is a typical two-state system treatment between two uncoupled levels $\{|L\uparrow\rangle, |R\uparrow\rangle\}$ and $\{|L\downarrow\rangle, |R\downarrow\rangle\}$ and therefore manifests as an avoided crossing between the bare states with eigen-energies given by 
\begin{equation}
    E^{ECT}_\pm = \frac{(\epsilon_L + \epsilon_R)}{2} \pm \frac{1}{2}\sqrt{(\epsilon_L - \epsilon_R)^2 + \Gamma_{ECT}^2}.
\end{equation}
At the CAR point $\epsilon_L = -\epsilon_R$, the expectation values $\langle d_{L\sigma}^\dagger d_{R\sigma} \rangle = \langle d_{R\sigma}^\dagger d_{L\sigma}\rangle = |L\sigma \rangle\langle R\sigma| = |R\sigma \rangle\langle L\sigma| = 0$, thereby restricting transitions between the singly-occupied states. The effective Hamiltonian is given by 
\begin{equation}
    H_{eff} = \sum_{\sigma = \uparrow \downarrow} \epsilon_L d_{L\sigma}^\dagger d_{L\sigma} + \epsilon_R d_{R\sigma}^\dagger d_{R\sigma} +  \frac{\Gamma_{CAR}}{\sqrt{2}} (d_{L\uparrow}^\dagger d_{R\downarrow}^\dagger - d_{L\downarrow}^\dagger d_{R\uparrow}^\dagger ) + h.c.,
\end{equation}
which can be mapped to a typical two-state Hamiltonian as well, with $|0\rangle$ and $|S\rangle$ as the levels and an avoided-crossing denoting the hybridization between the levels. The eigen-energies are given by 
\begin{equation}
    E^{CAR}_\pm = \frac{(\epsilon_L - \epsilon_R)}{2} \pm \frac{1}{2}\sqrt{(\epsilon_L + \epsilon_R)^2 + \Gamma_{CAR}^2}.
\end{equation}
At the centre of the two avoided-crossings, the energy difference between the branches is given by $\Gamma_{ECT}$ and $\Gamma_{CAR}$ respectively. From \cite{flindt-dynamic, dvir-eff-hamil} $\Gamma_{ECT} = \frac{\mu}{\mu^2 + \Delta^2}$ and $\Gamma_{CAR} = \frac{\Delta}{\mu^2 + \Delta^2}$. Therefore, the avoided-crossing gap is a decreasing function in the case of ECT and a small-valued increasing function in the case of CAR.

\end{document}